\documentclass [12pt]{article}
\begin{document}
\title{\textbf{Reliability of Module Based Software System  }}\author{ Rudrani Banerjee and Angshuman Sarkar\thanks{
Email:sangshu\_2000@yahoo.com} \\
\textbf{ \textit{Department of Statistics, Visva-Bharati
University, India}}} \maketitle
\begin{abstract}
This paper consider the problem of determining the reliability of
a software system which can be decomposed in a number of modules.
We have derived the expression of the reliability of a system
using the Markovian model for the transfer of control between
modules in order. We have given the expression of reliability by 
considering both benign and catastrophic failure. The expression of 
reliability presented in this work is applicable for some control software which
are designed to detect its own internal errors.
\end{abstract}
\section{Introduction}
Now a days large scale software systems are used in every walk of
life. The price of software are much higher than the cost of
hardware when we consider a huge computer intensive system.
Moreover the penalty cost incurred by a false outcome of a system
is enormous. To address such a challenge posed by this
technological trend, during the last three decades extensive
research has focused on the area of software reliability. The
consideration of software reliability is increasing because of the
growing emphasis on software that is reusable (as opposed to
software that is written for a terminal mission), where it is
essential to demonstrate that the system will perform reliably for
a variety of end-user applications.\\\indent A software system is
defined here as a " collection of programs and system files such
that the system files are accessed and altered only by the
programs in the collection ". Each element in this collection will
be called a module - for instance, a module might be a program, a
subprogram, or a file. The performance ( and hence the reliability
) of the system clearly depends on that of each individual module
and the relationship between these modules and the system; in this
regard a software system is quite similar to any other system.
However, the actual relationship between system and module
reliabilities is quite unique and depends on the specific
definition of software reliability as well as on the structure of
the overall system. In this paper we focus on software systems
that can be decomposed into a finite number of modules.\\\indent
In testing a software one may test the system as a whole, but in
practice, different organizational entities are assigned
responsibility of developing different modules. So it will be more
beneficial in the context of both cost and time test the
individual modules instead of testing them together. In order to
do this, some mathematical models, often referred to as Software
Reliability Growth Models (SRGM) are used to enable the software
reliability practitioners to estimate the expected future
reliability of a software under development and accordingly
allocate time, money, human resources to a project. Often these
reliability growth models express software system reliability in
terms of the individual module reliabilities which is favorable
under both time and cost constraints.\\\indent Cheung (2), first
expressed the system reliability in terms of the component
reliabilities. Poore et al. (1) suggest allocating the targeted
system reliability goal among the components and then testing the
individual components to verify whether the component
reliabilities meet the allocated goals at a specified level of
confidence, where as Easterling, Mazumdar, Spencer and Diegert
(6), has discussed this method may lead to estimates of overly
conservative sample size requirements for component testing. Yang
et. al. has implemented the idea of using testability to estimate
software reliability. They have also provided the basic steps
involve estimating testability, evaluating how well software was
written, and assessing the relationship between testing and usage
by assuming the modules are independently functioning.
 They have also compared their results with those obtained by using two reliability growth
models. Rajgopal et. al. has used a Markovian model for the
transfer of control between modules in order to develop the system
reliability expression in terms of the module reliabilities in
case of a dependent setup. They have also discussed a procedure
for determining the minimum number of tests required of each
module such that the probability of certifying a system whose
reliability falls below a specified value $R_0$ is less than a
specified small fraction $\beta$. Bondavalli et. al. has
considered the concept of benign failure and catastrophic failure
for determining the software reliability for a iterative program.
\\\indent In this paper we have expressed the system reliability in terms of testability of a
particular module following Yang et. al. for dependent modules and
have introduced the concepts of benign and catastrophic failure
following Bondavalli et. al. in case of a system where it can be
decomposed in a finite number of dependently functional modules.
The section 2 discuss the notations and preliminaries, section 3
gives the expression of the probability of correct output for a
specific input. Recent research [26] has shown a strong
correlation between reliability and coverage criteria (Lott et al.
(2005), Khun et. al. (2002), Yilmaz et. al. (2004) etc.), although
it is very difficult to quantify this relation. Dalal et al. [6]
and many more has examined this relationship between unit-test
statement coverage and system-test faults later attributed to
those units.
\\\indent Present work has been organized in 4 sections the section 2 gives the notation and preliminaries of
software reliability in terms testability of a module. In the 3rd
sections we have derived the probability of correct output of a
particular system corresponding to a particular input considering
both the case presence and absence of benign failure. In Section 4
we present a brief discussions about the procedure mentioned here.
\section{Notations and Preliminaries}
There is no rigorous definition of 'Quality'. But it can be weakly
defined as the fitness of purpose of any product to its users.
Similarly software quality is defined as the conformance to
explicitly stated functions and performance requirements,
explicitly documented development standards and implicit
characteristics that are expected of all professionally crafted
software(Cai Kai-Yuan Cai (3)). Alternatively, the quality of a
software may be characterized by some quality factors of a
software - reliability, efficiency, correctness, usability,
testability etc.\\\indent
 Reliability of a software system  may be
viewed as the expected value of probability of failure-free
operation of a program for a randomly chosen set of input
variables. The term failure in the context of software reliability
implies a result other than what was expected from the software
for a set of inputs. Following Voas et. al. (1995) we define the
testability of a particular system as the probability of failure
of the system for a particular input when it is assumed that there
is at least one fault in the system. Suppose we have a software
system which can be decomposed in $N$ modules. Thus the
testability of a particular module, say $i$th $(\forall i=1(1)N)$
module, is given by
\begin{eqnarray}
p_i&=&\mbox{Prob[ that the $i$th module will give incorrect output
$\mid$ there is at least one fault,}\nonumber\\& &
\mbox{probability distribution of input]}
\end{eqnarray}
The expression for the probability that the $i$th module will
contain error if the module has tested $n_i$ times successfully,
is given by the following (Yang et. al. (1998))
\begin{eqnarray}
\alpha_i(t)=\frac{\alpha_i(0)(1-p_i)^{n_i}}{\alpha_i(0)(1-p_i)^{n_i}+1-\alpha_i(0)}
\end{eqnarray}
where $\alpha_i(0)$ is the probability of failure of the system
before testing. Let $\pi_t(x)$ is the probability of a system
giving correct output corresponding to a particular set of input
$x$. The expression of $\pi_t(x)$ by assuming the independent
setup is given by (Yang et. al. (1998))
\begin{eqnarray}
\pi_t(x)=\prod_{i\in S}(1-q_i\alpha_i(t))
\end{eqnarray}
where $q_i$ is the revealibility of the $i$ th module and $S(x)$
is the set of those modules which will be executed by the input
$x$. The reliability of a software system is given by
\begin{eqnarray}
R_t=\int_{x\in X}\pi_t(x)\phi(x)dx
\end{eqnarray}
where $X$ is the set of all possible inputs and $\phi(x)$ is the
probability distribution of $x$.
\section{Detailed Expression of $\pi_t(x)$ for Dependent Setup}
A software system is necessarily an iterative. In each iteration a
particular module accepts a value and produce an output. The
outcomes of an individual iteration may be: i) success, i.e., the
delivery of a correct result, ii) a benign failure of the program,
i.e., an output that is not correct but does not, by itself, cause
the entire mission of the controlled system to fail, or iii) a
catastrophic failure, i.e., an output that causes the immediate
failure of the entire mission. The characterization of failures in
benign and catastrophic is discussed with example by Bondavalli.
et. al. ().  In this section we derive the expression of
$\pi_t(x)$ first of all only considering the catastrophic failure
and then in the subsequent subsection considering the benign and
catastrophic failure simultaneously.
\subsection{Expression of $\pi_t(x)$: No Benign Failure in the System}
Consider the above software system with $N$ modules. Let $p_{ij}$
be the probability that the control from the $i$th module will be
transferred to the $j$th module with correct execution $(\forall
i=1(1)N,\forall j=1(1)N)$. Let $S$ be a state of successful
completion of the system. As $S$ is achievable from any one of the
module so we define $p_{iS}$ $(\forall i=1(1)N)$ as the
probability of successful completion of the mission from the $i$th
module. Here we must have $p_{iS}+\sum_{j=1}^np_{ij}=1$.\\\indent
As we have a faulty system, that is, we have a system where there
is at least one fault or if the faults can be classified into
categories then there are at most one fault of each category. So
we introduce another state $F$, i.e.,  unsuccessful completion of
the mission. As any module may be faulty so the state $F$ also can
be achieved from any of the module. We define $p_{iF}$ as the
probability of unsuccessful completion of the module $i$ $(\forall
i=1(1)N)$. The transition probability matrix takes the following
form for the above setup.
\begin{eqnarray}
Q=\left(%
\begin{array}{cccccc}
  p_{11}(1-\alpha^x_1(t)) & p_{12}(1-\alpha^x_1(t)) & ... & p_{1N}(1-\alpha^x_1(t)) & p_{1S}(1-\alpha^x_1(t)) & \alpha^x_1(t) \\
  p_{21}(1-\alpha^x_2(t)) &p_{22}(1-\alpha^x_2(t)) & ... & p_{2N}(1-\alpha^x_2(t)) & p_{2S}(1-\alpha^x_2(t)) & \alpha^x_2(t) \\
  ... & ... & ... & ... & ... & ... \\
  p_{N1}(1-\alpha^x_N(t)) &p_{N2}(1-\alpha^x_N(t)) & ... & p_{NN}(1-\alpha^x_N(t)) & p_{NS}(1-\alpha^x_N(t)) & \alpha^x_N(t) \\
  0 & 0 & ... & 0 & 1 & 0 \\
  0 & 0 &... & 0 & 0 & 1 \\
\end{array}%
\right)
\end{eqnarray}
where $\alpha^x_i(t)$ is the probability of faulty completion of
the $i$th module for the input x. The expression of
$\alpha^x_i(t)$ is given by
\begin{eqnarray}\alpha^x_i(t)=q_i\alpha_i(t) \end{eqnarray}If we
assume that the first block is the control block then the
probability of correct completion of the mission for the given
input $x$ is given by (Parzen (1962))
\begin{eqnarray}
\pi_t(x)=\sum_{i=1}^N(I_N-\hat{Q})^{-1}_{1i}p_{iS}(1-\alpha^x_i(t))
\end{eqnarray}
where $\hat{Q}$ is the sub-matrix of $Q$ deleting its last two
columns and rows.
\subsection{Expression of $\pi_t(x)$: Benign Failure and Catastrophic Failure are in the System}
From the software viewpoint solely, and without referring to any
specific application, we assume here that all detected failures
(default safe values of the control outputs from the computer) do
not prevent the mission to continue and are in this sense benign,
whereas undetected failures are conservatively assumed to have a
"catastrophic" effect on the controlled system. Obviously, if
knowledge of the consequences of software failures on the system
was available for a specific system, the proper splitting of
software failures into benign and catastrophic could be precisely
made. We make the following assumption to model the system.
\\\indent Suppose $SS$ is a state where the total system, that is all the $N$ modules, runs without
any fault of either kind. Let $B_i$ be the state where the system
is running in benign failure of $i$th level, that is after $i$
iterations the system will enter in the state $SS$. As the
previous subsection $S$ and $F$ denotes the successful completion
of the mission and completion of the mission with a failure
respectively. The mission will fail if their is a catastrophic
failure in the system. Let us also assume that if there is a
benign failure of length greater than a threshold value, say
$n_c$, then the system will enter in a catastrophic failure
region. Although this assumption will take the model a little away
from reality, a model should be good enough to handle a benign
failure of any arbitrary random length, but this assumption will
make the calculation of reliability expression easier which will
increase its practical application. At this point note that the
state $S$, that is the successful completion of the program, can
be achieved only from the state $SS$, where as the state $F$ can
be achieved from any of the state $SS$ or $B_i$'s $(\forall
i=1(1)N)$, but we assume here the control will be transferred from
the state $B_i$ to $B_{i-1}$ only to reduce the number of
parameters in the model.\\\indent The transition probability
matrix will be as follows
\begin{eqnarray}
Q=\left(%
\begin{array}{ccccccccc}
  Q_{00} & Q^b_{01} & Q^b_{02} & ...&Q^b_{0(n_c-2)}&Q^b_{0(n_c-1)} & Q^b_{0n_c} & S^0&F^0 \\
 Q^b_{10} & O & O & ...&O&O & O & \bar{0}&\bar{0} \\
O & Q^b_{21} & O & ... &O&O& O & \bar{0}&\bar{0}\\
...& ... & ...& ... & ... & ...&...&...&... \\
  O & O & O & ... & Q^b_{(n_c-1)(n_c-2)}&O&O & \bar{0}&\bar{0} \\
  O & O & O & ... &O& Q^b_{n_c(n_c-1)}&O & \bar{0}&\bar{0} \\
  \bar{0}'& \bar{0}'&\bar{0}'&...&\bar{0}'&\bar{0}'&\bar{0}'&1&0\\
  \bar{0}'& \bar{0}'&\bar{0}'&...&\bar{0}'&\bar{0}'&\bar{0}'&0&1\\
\end{array}%
\right)
\end{eqnarray}
Here the matrix $Q_{00}$ is a $N\times N$ matrix which describes
that the flow is running without entering in benign failure or
catastrophic failure. The matrix $Q^b_{0k}$ is also a $N\times N$
matrix giving the transition probabilities of the flow of control
from stable state to the $k$th level benign failure $(\forall
k=1(1)n_c)$. Similarly, the matrix $Q^b_{kl}$ which is also
$N\times N$ denotes the transition probabilities of the control
entering from the $k$th level benign failure to $l$ th level
$(\forall k=1(1)n_c\forall l=1(1)n_c)$. From the $k$th level
benign failure we can only achieve the $k-1$th level benign
failure so $Q^b_{kl}=O$ ($\forall l\neq k-1$). Where $O$ is the
null matrix of order $N\times N$. $S^0$ is a $N\times 1$ vector of
the transition probabilities of successful completion of the
mission from the stable state. As the mission can terminate
successfully only from the stable state so the rest of the entries
in this column are all zero. $\bar{0}$ denotes a null vector of
length $N$ and $\bar{0}'$ denotes transpose of $\bar{0}$. Finally,
$F^0$ is a column vector of length $N$ giving probabilities of
reaching the state of catastrophic failure from the stable state.
\\\indent To give the structure of sub-matrices $Q_{00}$, let us
define $p^{SS}_{ij}$ be the probability of the control to enter
from the $i$th module to $j$th module in the state $SS$. So the
matrix $Q_{00}$ is given by
\begin{eqnarray}
Q_{00}=\left(%
\begin{array}{cccc}
  p^{SS}_{11} & p^{SS}_{12} & ... & p^{SS}_{1N} \\
  p^{SS}_{21} & p^{SS}_{22} & ... & p^{SS}_{2N} \\
  ... & ... & ... & ... \\
  p^{SS}_{N1} & p^{SS}_{N2} & ... & p^{SS}_{NN} \\
\end{array}%
\right)
\end{eqnarray}
\\\indent Let us also define $p^{SB}_{ij}$ be the probability that
the control will be transferred from the module $i$ to the module
$j$ from the state $SS$ to any of benign failure. Let also $p^B_k$
the probability that the control will enter in $B_k$, thus the
probability that the control will enter in the $j$th module from
the $i$th module in the state $B_k$ is given by
$p^{SB}_{ij}p^B_k$. So the matrix $Q^b_{0k}$ will take the
following form
\begin{eqnarray}
Q^b_{0k}=\left(%
\begin{array}{cccc}
  p^{SB}_{11}p^B_k & p^{SB}_{12}p^B_k & ... & p^{SB}_{1N}p^B_k \\
  p^{SB}_{21}p^B_k & p^{SB}_{22}p^B_k & ... & p^{SB}_{2N}p^B_k \\
  ... & ... & ... & ... \\
  p^{SB}_{N1}p^B_k & p^{SB}_{N2}p^B_k & ... & p^{SB}_{NN}p^B_k \\
\end{array}%
\right)
\end{eqnarray}
If $p_{iS}$ and $p_{iF}$ is respectively the successful completion
of the mission and achieving catastrophic failure from the $i$th
module. Then we must have
\begin{eqnarray}
\sum_{j=1}^Np^{SS}_{ij}+\sum_{k=1}^{n_c}p^B_k\sum_{j=1}^Np^{SB}_{ij}+p_{iS}+p_{iF}=1\hspace{.3in}\forall
i=1(1)N
\end{eqnarray}
The matrix $Q^b_{kk-1}$ takes the following form
\begin{eqnarray}
Q^b_{kk-1}=\left(%
\begin{array}{cccc}
  p^{bb}_{11} & p^{bb}_{12} & ... & p^{bb}_{1N} \\
  p^{bb}_{21} & p^{bb}_{22} & ... & p^{bb}_{2N}\\
  ... & ... & ... & ... \\
  p^{bb}_{N1} & p^{bb}_{N2} & ... & p^{bb}_{NN} \\
\end{array}%
\right)
\end{eqnarray}
Here we have
\begin{eqnarray}
\sum_{j=1}^Np^{bb}_{ij}=1\hspace{.3in}\forall i=1(1)N
\end{eqnarray}
Finally, the matrix $Q^b_{10}$ is the matrix of transition
probabilities, say $p^{bS}_{ij}$, that the flow of control will be
transferred from the $i$th to the $j$th module and from the $B_1$
to $SS$. Here also
\begin{eqnarray}
\sum_{j=1}^Np^{bS}_{ij}=1\hspace{.3in}\forall i=1(1)N
\end{eqnarray}
By assuming as before the first module as the control module the
expression of $\pi_t(x)$ is given
\begin{eqnarray}
\pi_t(x)=\sum_{i=1}^N(I_{Nn_c}-\hat{Q})^{-1}_{1i}p_{iS}
\end{eqnarray}
where $\hat{Q}$ is once again the sub-matrix of $Q$ deleting its
last two columns and rows.
\section{Conclusions}
In this work we have given an expression of the reliability of a
software system which can be divided in a finite number of
modules. The transition probabilities we have considered can be
easily estimated using maximum likelihood method of estimation.
\\\indent Consider the setup without benign failure,
suppose $i$th block is tested $n_i$ times, out of which $x^i_j$
times the control is transferred to the $j$th state $(\forall
i=1(1)N \& \forall j=1(1)N,S,F)$. The maximum likelihood estimates
of $p_{ij}$ is $x^i_j/(\sum_{i=1}^N x^i_j+x^i_S)$ and that of
$\alpha^x_i(t)$ is $x^i_F/n_i$. Hence estimate of $\pi_t(x)$ can
be obtained and let it be denoted by $\hat{\pi}_t(x)$. Finally the
estimate of reliability of a system can be given by
\begin{eqnarray}
\hat{R_t}=\frac{1}{|W|}\sum_{x\in W}\hat{\pi}_t(x)
\end{eqnarray}
where $W$ is the set of all inputs which are used for testing.
This is an extension of some previous work and the model what we
have considered are more realistic for some control software which
are designed to detect its own internal errors and then issue a
safe output and reset itself to a known state from which the
program is likely to proceed correctly.


\begin{thebibliography}{99}
\bibitem{Bertolino and Strigini(1996)} A.Bertolino and L.Strigini (1996). On The Use of Testability Measures for Dependability Assessment.
IEEE Trans on Software Engineering 22(2):97-108. 
\bibitem{Beightler and Phillips(1976)}C. Beightler and D.T. Phillips (1976). Applied Geometric Programming. John Wiley
\& Sons,Inc., New York. 
\bibitem{Cinlar(1975)} E.Cinlar (1975). Introduction to Stochastic Processes. Prentice-Hall, Inc., Englewood
Clios, N.J. 
\bibitem{E.C.Soistman and K.B.Ragsdale(1984)}E.C.Soistman and K.B.Ragsdale (1984). Combined Hardware/Software Reliability
Prediction Methodology. Rome Air Development Center Contract Report
OR. 18 - 173, Vol.2.
\bibitem{Parzen 1962} E.Parzen. (1962), Stochastic Processes, Holden-Day, San Francisco, Calif.
\bibitem{Poore et. al.(1993)} J.H.Poore, H.D.Mills and D.Mutchler (1993). Planning and Certifying Software
System Reliability. IEEE Software, 88-99. 
\bibitem{Voas et. al. 1995} J.M.Voas and K.W.Miller (1995). Software Testability: The new verification. IEEE Software.
17-28.
\bibitem{J.Rajgopal and D.L.Bricker(1995)}J.Rajgopal and D.L.Bricker (1995). An Algorithm for Solving The Polynomial GP
Problem, Based on Generalized Programming. Department of Industrial
Engineering, University of Pittsburgh, Technical Report No.TR95-10.
\bibitem{J.Rajgopal and M.Mazumdar (1995)} J.Rajgopal and M.Mazumdar (1995). Designing Component Test Plans for Series
System Reliability via Mathematical
Programming. Technometrics 37,195-212.
\bibitem{J.Rajgopal and
M.Mazumdar (1997)} J.Rajgopal and
M.Mazumdar (1997). Minimum Cost Component Test Plans for evaluating
Reliability of a Highly Reliable Parallel System. Naval Research
Logistics 44, 401-418.
\bibitem{K.S.Al-Sultan, M.F.Hussain, J.S.Nizami}K.S.Al-Sultan, M.F.Hussain and J.S.Nizami (1996). A Genetic Algorithm for The Set Covering
Problem. The Journal of the Operational Research Soc. 47,
5, 702-709.
\bibitem{K.Siegrist(1998)}K.Siegrist(1998). Reliability of Systems with Markov Transfer of Control. IEEE
Transaction on Software Engineering 14, 1049-1053.
\bibitem{Miller et. al. 1992}K.W.Miller, L.J.Morrel, R.E.Noonan, S.K.Park, D.M.Nicol, B.M.Murril and J.M.Voas (1992)
Estimating the probability of failure when testing reveals no
failure. IEEE Trans. on Software Engineering 18(1): 33-43.
\bibitem{Yang et.al. 1998}Mark C.K. Yang, W.Eric Wong, Alberto Pasquini (1998). Applying Testability to Reliability
Estimation, Proc. of IEEE International Symposium on Software
Reliability Engineering, Puderborn 90-99. 
\bibitem{M.Avriel,R.S.Dembo and U.Plassy(1975)} M.Avriel,R.S.Dembo and U.Plassy (1975). Solution of Generalized
Geometric Programs. International Journal for Numerical methods in
Engineering 9, 149-168.
\bibitem{Cheung 1980}R.C.Cheung (1980). A User-Oriented Reliability Model. IEEE Trans. Software Engineering, SE-
6(2): 118-125.
\bibitem{Easterling(1991)}R.C.Easterling, M.Mazumdar, F.W.Spencer and K.V.Diegert (1991). System Based Component
Test Plan and Operating Characteristics: Binomial Data
Technometrics 33,287- 298.
\bibitem{Gal et. al. 1974}S.Gal (1974). Optimal Test Design for Reliability Demonstration. Operational Research
22, 1236-1242.
\bibitem{S.Ghosh, A.P.Mathur, J.R.Horgan, J.J.Li, W.E.Wong (1997)}S.Ghosh, A.P.Mathur, J.R.Horgan, J.J.Li, W.E.Wong(1997). Software Fault Injection
Testing on a Distributed System - A Case Study, Proc. of the 1st
International Quality Week Europe, Brussels, Belgium.
\bibitem{S.Wolfram(1996)} S.Wolfram(1996). The Mathematics Book (3rd edition). Cambridge University Press
and Wolfram Media, Inc.Champaign, 3.
\bibitem{W.Kuo (1992)}W.Kuo (1992). Software Reliability. Maynards Industrial Engineering Handbook, 4th
edition(W.K.Hodson, Editor-in-Chief), 11116-11122.
\bibitem{W.Eric Wong, J.R.Horgan, S.London and Aditya P.Mathur (1998)}W.Eric Wong, J.R.Horgan, S.London and Aditya P.Mathur (1998). Effect of test set minimization
on Fault Detection Effectiveness. Software-Practice and
Experience, 28(4): 347-369.
\end{thebibliography}
\end{document}